\documentclass[pra,aps,preprint,amsmath,amssymb,showpacs]{revtex4}
\usepackage{graphicx}
\usepackage{epsfig}
\include{graphics}
\usepackage{epstopdf} 

\begin{document}

\title{Robust transmission stabilization and dynamic switching 
in broadband hybrid waveguide systems with nonlinear gain and loss}
\author{Quan M. Nguyen$^{1}$, Avner Peleg$^{2}$, and Thinh P. Tran$^{3}$}
\affiliation{$^{1}$Department of Mathematics, International University, 
Vietnam National University-HCMC, Ho Chi Minh City, Vietnam}
\affiliation{$^{2}$Department of Mathematics, State University of New York
at Buffalo, Buffalo, New York 14260, USA}
\affiliation{$^{3}$Department of Theoretical Physics, University of Science, 
Vietnam National University-HCMC, Ho Chi Minh City, Vietnam}

\date{\today}

\begin{abstract} We develop a method for transmission stabilization 
and robust dynamic switching for colliding optical soliton 
sequences in broadband waveguide systems with nonlinear gain and loss. 
The method is based on employing hybrid waveguides,  
consisting of spans with linear gain and cubic loss, and spans 
with linear loss, cubic gain, and quintic loss. 
We show that amplitude dynamics is described by a hybrid 
Lotka-Volterra (LV) model, and use the model to determine 
the physical parameter values 
required for enhanced transmission stabilization and switching.  
Numerical simulations with the coupled nonlinear 
Schr\"odinger equations confirm the predictions of the LV model,  
and show complete suppression of radiative instability,  
which enables stable transmission over 
distances larger by an order of magnitude 
compared with uniform  waveguides with linear gain and cubic loss. 
Moreover, multiple on-off and off-on 
dynamic switching events are demonstrated over a wide 
range of soliton amplitudes, showing the superiority 
of hybrid waveguides compared with static 
switching in uniform waveguides.               
\end{abstract}

\maketitle

\section{Introduction}
\label{Introduction}
Recent years have seen a dramatic increase in research on 
broadband optical waveguide systems
\cite{Agrawal2001,Mollenauer2006,Gnauck2008,Essiambre2010}. 
This increase in research 
efforts is driven by a wide range of applications, which include 
increasing transmission rates in fiber optics 
communication systems \cite{Mollenauer2006,Gnauck2008,Essiambre2010}, 
enhancing data processing and transfer on computer chips 
\cite{Soref2006,Dekker2007,Agrawal2007a,Gaeta2008}, 
and enabling multiwavelength optical waveguide lasers
\cite{Chow96,Shi97,Sun2002,Zhang2009,Oh2009,Liu2013}. 
Transmission in broadband systems is often based on 
wavelength-division-multiplexing (WDM), where many pulse sequences 
propagate through the same waveguide. The pulses in each sequence 
(each ``frequency channel'') propagate with the same group velocity, 
but the group velocity differs for pulses from different sequences. 
As a result, intersequence pulse collisions are very frequent, 
and can lead to severe transmission degradation
\cite{Agrawal2001,Mollenauer2006,Tkach97,Iannone98,Essiambre2010}. 
On the other hand, the significant collision-induced effects 
can be used for controlling 
the propagation, for tuning of optical pulse parameters, such as energy, 
frequency, and phase, and for transmission switching, 
i.e., the turning on or off of transmission of one or 
more of the pulse sequences \cite{PNC2010,PC2012,CPJ2013}.

One of the most important processes affecting pulse propagation 
in nonlinear waveguide systems is due to nonlinear loss or gain. 
Nonlinear loss (gain) can arise in optical waveguides due to 
multiphoton absorption (emission) or due to gain (loss) 
saturation \cite{Boyd2008,Prasad2008}. 
For example, cubic loss due to two-photon absorption (TPA) 
plays a key role in pulse dynamics in a variety of waveguides, 
including silicon waveguides 
\cite{Soref2006,Agrawal2007a,Dekker2007,Gaeta2008,Malomed89,Stegeman89,
Silberberg90,Aceves92,Prasad95,Kivshar95,Perry97,Agrawal2007b,
Skryabin2008,Prasad2008b,Gaeta2012}. 
Furthermore, cubic gain and quintic loss 
are essential parts of the widely used Ginzburg-Landau (GL) 
model for pulse dynamics in mode-locked lasers 
\cite{Moores93,Moores95,Akhmediev96,Kramer2002,Kutz2006,Wise2010}.    
The main effect of nonlinear loss (gain) on single pulse 
propagation is a continuous decrease (increase)  
of the pulse amplitude, which is qualitatively similar 
to the one due to linear loss (gain) \cite{Malomed89}. 
Nonlinear loss (gain) also strongly affects optical pulse collisions, 
by causing an additional decrease (increase) of pulse 
amplitudes \cite{PNC2010,PC2012,CPJ2013,PNG2014}. 
This collision-induced amplitude shift, which is commonly known 
as interchannel crosstalk, can be a major impairment in broadband 
nonlinear waveguide systems. For example, recent experiments 
have shown that crosstalk induced by cubic loss (due to TPA) 
plays a key role in silicon nanowaveguide 
WDM systems \cite{Gaeta2012}. 
More specifically, the experiments demonstrated that  
TPA-induced crosstalk can lead to relatively 
high values of the bit-error-rate even in a 
WDM system with two channels \cite{Gaeta2012}.    
Thus, it is important to find ways to suppress 
the detrimental effects of nonlinear gain-loss crosstalk.

In several recent studies \cite{PNC2010,PC2012,CPJ2013} 
we provided a partial solution to this key problem and 
to an equally important challenge concerning the possibility 
to use the nonlinear crosstalk for broadband transmission 
switching. Our approach was based on showing that 
amplitude dynamics of N sequences of colliding optical solitons 
can be described by Lotka-Volterra (LV) models for N species, 
where the exact form of the LV model depends on the 
nature of the waveguide's gain-loss profile  \cite{PNC2010,PC2012}. 
Stability analysis of the steady states of the LV models  
was used to guide a clever choice of linear amplifier gain, 
which in turn leads to transmission stabilization, i.e., 
the amplitudes of the propagating pulses approach 
desired predetermined values \cite{PNC2010,PC2012,CPJ2013}. 
Furthermore, in Ref. \cite{CPJ2013}, we showed that static 
on-off and off-on transmission switching can be realized 
by an abrupt change in the waveguide's nonlinear gain or loss 
coefficients. The design of the switching setups reported in 
Ref. \cite{CPJ2013} was also guided by linear stability 
analysis of the steady states of the LV model.

The results of Refs. \cite{PNC2010,PC2012,CPJ2013} demonstrate 
the potential of employing crosstalk induced by nonlinear loss or gain  
for transmission control, stabilization, and switching. However, 
these results are still quite limited due to the following 
reasons. First, despite the progress made in 
Refs. \cite{PNC2010,PC2012,CPJ2013}, the problem of robust 
transmission stabilization is still unresolved. 
In particular, for uniform waveguides with linear gain and cubic loss, 
such as silicon waveguides, radiative instability due to the growth 
of small amplitude waves is observed already at a distance $z\simeq 200$ 
even for cubic loss coefficient values as small as $0.01$ \cite{PNC2010}. 
The radiative instability can be partially mitigated by employing 
uniform waveguides with linear loss, cubic gain, and quintic loss,  
i.e., waveguides with a GL gain-loss profile \cite{PC2012,CPJ2013}. 
However, this uniform GL gain-loss setup is also limited, 
since the initial soliton amplitudes need to be close 
to the steady state values for transmission stabilization 
to be achieved. Second, the switching 
setup studied in Ref. \cite{CPJ2013} 
is also quite limited, since it is based on a {\it static} 
change in the waveguide's nonlinear gain-loss coefficients. 
Moreover, only one switching event was demonstrated 
in this study, and off-on transmission was restricted to 
amplitude values larger than 0.65. In view of the limitations 
of these uniform waveguide setups, it is important to look 
for more robust ways for realizing stable long-distance 
propagation and broadband transmission switching.

In the current paper we take this important task, 
by developing a method for transmission stabilization 
and switching in broadband waveguide systems, 
which is based on employing hybrid waveguides with 
a clever choice of the physical parameters. 
The hybrid waveguides consist of odd-numbered spans 
with linear gain and cubic loss, and even-numbered 
spans with a GL gain-loss profile.   
Transmission switching is {\it dynamically} realized 
by fast changes in linear amplifier gain. The robustness of 
the approach is demonstrated for two sequences of 
colliding optical solitons. 
We show that the dynamics of soliton amplitudes is described 
by a hybrid LV model. We then use stability analysis 
for the steady states of the LV model to determine the 
physical parameters that lead to suppression of radiative 
instability and as a result, to a drastic enhancement in 
transmission stability and switching robustness.     
The hybrid nature of the waveguides and the corresponding 
LV model plays a key role in the improvement. 
The predictions of the hybrid LV model are confirmed 
by numerical simulations with the full system of 
coupled nonlinear Schr\"odinger (NLS) equations. 
The results of the latter simulations show complete suppression of radiative instability, which enables 
stable propagation over distances larger by an order of magnitude compared with 
the results reported in Ref. \cite{PNC2010} for uniform waveguides 
with linear gain and cubic loss. 
Moreover, multiple dynamic on-off and off-on  
switching events are demonstrated over a significantly  
wider range of soliton amplitudes compared with 
that reported in Ref. \cite{CPJ2013} for a single static  
switching event in uniform waveguides with a GL 
gain-loss profile. The increased robustness of 
off-on switching in hybrid waveguides 
can be used for transmission recovery, 
that is, for the stable amplification of optical pulse 
sequences that experienced significant energy decay.

We choose optical solitons as an example for 
the propagating pulses for the following reasons. 
First, in many broadband optical systems the waveguides are 
nonlinear and pulse propagation is accurately described by 
a perturbed NLS equation \cite{Silberberg90,Aceves92,Kivshar95,
Agrawal2007a,Dekker2007,Gaeta2008}. 
Furthermore, optical soliton generation and propagation 
in the presence of two-photon and three-photon absorption  
was experimentally demonstrated in a variety of waveguide setups 
\cite{Silberberg90b,Agrawal2007b,Silberberg2008,
Skryabin2008,Fan2010,Husko2013}. 
Second, since the unperturbed NLS equation is an integrable 
model \cite{Zakharov84}, derivation of analytic results for the 
effects of nonlinear gain or loss on interpulse 
collisions can be done in a rigorous manner. 
Third, due to the soliton properties, soliton-based 
information transmission and processing in nonlinear 
broadband waveguide links is considered to be highly 
advantageous compared with other transmission 
methods \cite{Agrawal2001,Mollenauer2006,Iannone98}.

The rest of the paper is organized as follows. 
In Sec. \ref{Model}, we present the coupled-NLS model 
for pulse propagation in hybrid waveguides, 
along with the corresponding  hybrid LV model for 
amplitude dynamics. We then use stability 
analysis of the equilibrium states of the  
hybrid LV model to obtain the physical 
parameter values required for robust transmission stabilization 
and broadband switching. In Sec. \ref{Simulations}, 
we present the results of numerical simulations 
with the coupled-NLS model for stable long-distance 
propagation and multiple transmission switching events.  
We also analyze these results in comparison   
with the predictions of the LV model. 
Section \ref{Conclusions} is reserved for conclusions.


\section{Coupled-NLS and Lotka-Volterra models for pulse propagation}
\label{Model}
We consider two sequences of optical solitons propagating with different 
group velocities in a hybrid waveguide system, in which the gain-loss profile 
is different for different waveguide spans. 
We take into account second-order dispersion, Kerr nonlinearity, 
as well as linear and nonlinear gain and loss. 
We denote by $z$ distance along the waveguide, 
and assume that the gain-loss profile consists of 
linear gain and cubic loss in odd-numbered spans 
$z_{2m} \le z <z_{2m+1}$, and of linear loss, cubic gain, 
and quintic loss in even-numbered spans $z_{2m+1}\le z <z_{2m+2}$, 
where $0\le m \le M$, $M\ge 0$, and $z_{0}=0$. 
Thus, the propagation is described by the following 
system of coupled-NLS equations: 
\begin{eqnarray} &&
i\partial_z\psi_{j}+\partial_{t}^2\psi_{j}+2|\psi_{j}|^2\psi_{j}
+4|\psi_{k}|^2\psi_{j}=ig_{j}^{(l)}\psi _{j}/2+L_{l}(\psi _{j},\psi _{k}),
\label{Hybrid1}
\end{eqnarray} 
where $t$ is time, $\psi_{j}$ is the electric field's envelope for 
the $j$th sequence, $g_{j}^{(l)}$ is the linear gain-loss coefficient, 
and $L_{l}(\psi _{j},\psi _{k})$ describes nonlinear gain-loss effects. 
The indexes $j$ and $k$ run over pulse sequences, 
i.e., $j=1,2$, $k=1,2$, while $l$ runs over the two gain-loss profiles. 
The second term on the left hand side of Eq. (\ref{Hybrid1}) 
corresponds to second-order dispersion, while the third 
and fourth terms describe the effects of intrasequence and 
intersequence interaction due to Kerr nonlinearity.

The optical pulses in the $j$th sequence are 
fundamental solitons of the unperturbed NLS equation 
$i\partial_z\psi_{j}+\partial_{t}^2\psi_{j}+2|\psi_{j}|^2\psi_{j}=0$. 
The envelopes of these solitons are given by 
$\psi_{sj}(t,z)=\eta_{j}\exp(i\chi_{j})\mbox{sech}(x_{j})$,
where $x_{j}=\eta_{j}\left(t-y_{j}-2\beta_{j} z\right)$, 
$\chi_{j}=\alpha_{j}+\beta_{j}(t-y_{j})+
\left(\eta_{j}^2-\beta_{j}^{2}\right)z$, 
and $\eta_{j}$, $\beta_{j}$, $y_{j}$, and $\alpha_{j}$ 
are related to the soliton amplitude, group velocity (and frequency), 
position, and phase, respectively. 
We assume a large group velocity difference 
$|\beta_{1}-\beta_{2}|\gg 1$, so that the solitons undergo a large number 
of fast intersequence collisions. Due to the presence of nonlinear 
gain or loss the solitons experience additional changes in their 
amplitudes during the collisions, and this can be used for achieving 
robust transmission stabilization and switching.

The nonlinear gain-loss term $L_{1}(\psi _{j},\psi _{k})$ 
in odd-numbered spans is  
\begin{eqnarray} &&      
L_{1}(\psi_{j},\psi _{k})=
-i\epsilon _{3}^{(1)}\left|{\psi_{j}}\right|^2\psi_{j}
-2i\epsilon _3^{(1)}\left|{\psi_{k}}\right|^2\psi _{j},  
\label{Hybrid1_a}
\end{eqnarray} 
where $\epsilon_{3}^{(1)}$ is the cubic loss coefficient. 
The first and second terms on the right hand side of Eq. (\ref{Hybrid1_a}) 
describe intrasequence and intersequence interaction due to cubic loss.    
The nonlinear gain-loss term $L_{2}(\psi_{j},\psi_{k})$ 
in even-numbered spans is  
\begin{eqnarray} &&      
L_{2}(\psi_{j},\psi _{k})=
i\epsilon_{3}^{(2)}|\psi_{j}|^2\psi_{j}
+2i\epsilon_{3}^{(2)}|\psi_{k}|^2\psi_{j}
\nonumber \\&&
-i\epsilon_{5}|\psi_{j}|^4\psi_{j}-3i\epsilon_{5}|\psi_{k}|^4\psi_{j}
-6i\epsilon_{5}|\psi_{k}|^2|\psi_{j}|^2\psi_{j},  
\label{Hybrid1_b}
\end{eqnarray} 
where $\epsilon _3^{(2)}$ and $\epsilon_{5}$ are the cubic gain and quintic 
loss coefficients, respectively. The first and second terms on the right 
hand side of Eq. (\ref{Hybrid1_b}) describe intrasequence and intersequence 
interaction due to cubic gain, while the the third, fourth, and fifth 
terms are due to quintic loss effects.

In several earlier works, we showed that 
amplitude dynamics of $N$ colliding sequences of optical 
solitons  in the presence of linear and nonlinear gain or 
loss can be described by LV models for $N$ species, 
where the exact form of the model depends 
on the nature of the waveguide's 
gain-loss profile \cite{NP2010,PNC2010,PC2012}. 
The derivation of the LV models was based on the following 
assumptions. (1) The temporal separation $T$ between 
adjacent solitons in each sequence is a constant satisfying: 
$T \gg 1$. In addition, the amplitudes are equal for all solitons 
from the same sequence, but are not necessarily equal for solitons from 
different sequences. This setup corresponds, for example, 
to return-to-zero phase-shift-keyed soliton transmission.  
(2) The pulses circulate in a closed optical waveguide loop.  
(3) As $T\gg 1$, the pulses in each sequence are 
temporally well-separated. As a result, intrasequence 
interaction is exponentially small and is neglected.

Under the above assumptions, the soliton sequences are periodic, 
and as a result, the amplitudes of all pulses in a given sequence 
undergo the same dynamics. Consider first odd-numbered waveguide 
spans, where the gain-loss profile consists of linear gain and 
cubic loss. Taking into account collision-induced amplitude 
shifts due to cubic loss and single-pulse amplitude changes 
due to linear gain and cubic loss, we obtain the following 
equation for amplitude dynamics of $j$th sequence 
solitons \cite{PNC2010}: 
\begin{eqnarray} &&  
\frac{d\eta _{j}}{dz}=
\eta_{j}\left(
g_{j}^{(1)}-4\epsilon_{3}^{(1)}\eta_{j}^2/3
-8\epsilon_{3}^{(1)}\eta_{k}/T\right), 
\label{Hybrid3}
\end{eqnarray} 
where $j=1,2$ and $k=1,2$. In WDM transmission systems, it is often 
required to achieve a transmission steady state, in which pulse 
amplitudes in all sequences are nonzero constants.   
We therefore look for a steady state of Eq. (\ref{Hybrid3}) 
in the form $\eta^{(eq)}_{1}=a>0$, $\eta^{(eq)}_{2}=b>0$,     
where $a$ and $b$ are the desired equilibrium 
amplitude values. This requirement yields: 
$g_{1}^{(1)}=4\epsilon_3^{(1)}(a^{2}/3+2b/T)$ and 
$g_{2}^{(1)}=4\epsilon_3^{(1)}(b^{2}/3+2a/T)$. 
Note that in transmission stabilization and off-on switching 
we use $a=b=\eta$, corresponding to the desired situation 
of equal amplitudes in both sequences. 
In contrast, in on-off switching, we use $a \ne b$, 
since turning off of transmission of only one 
sequence is difficult to realize with $a=b$.   
Also note that switching is obtained by fast changes in the 
values of the $g_j^{(1)}$ coefficients, such that $(a,b)$ becomes 
asymptotically stable in off-on switching and unstable in on-off switching. 
The switching is realized {\it dynamically}, via appropriate 
fast changes in amplifier gain, and is thus very different 
from the {\it static} switching that was 
studied in Ref. \cite{CPJ2013}.

The LV model for amplitude dynamics in even-numbered spans is 
obtained by taking into account collision-induced amplitude 
shifts due to cubic gain and quintic loss, as well as 
single-pulse amplitude changes due to linear loss, cubic gain, 
and quintic loss. The derivation yields the following equation
for amplitude dynamics of the $j$th 
sequence solitons \cite{PC2012}: 
\begin{eqnarray} &&
\frac{d \eta_{j}}{dz}=
\eta_{j}\left[g_{j}^{(2)}+4\epsilon_{3}^{(2)}\eta_{j}^{2}/3
-16\epsilon_{5}\eta_{j}^{4}/15
+8\epsilon_{3}^{(2)}\eta_{k}/T
-8\epsilon_{5}\eta_{k}\left(2\eta_{j}^{2}+\eta_{k}^{2}\right)/T
\right].   
\label{Hybrid4}
\end{eqnarray} 
Requiring that $(\eta,\eta)$ is a steady state of Eq. (\ref{Hybrid4}), 
we obtain $g_{j}^{(2)}=4\epsilon_{5}\eta(-\kappa\eta/3+4\eta^{3}/15
-2\kappa/T+6\eta^{2}/T)$, where 
$\kappa=\epsilon_{3}^{(2)}/\epsilon_{5}$ and $\epsilon_{5}\ne 0$.
Note that in even-numbered spans, the value of $\kappa$ is used 
for further stabilization of transmission and switching. 


Transmission stabilization and switching are guided by stability 
analysis of the steady states of Eqs. (\ref{Hybrid3}) 
and (\ref{Hybrid4}). We therefore turn to describe the results 
of this analysis, starting with the LV model (\ref{Hybrid3}). 
We consider the equilibrium amplitude values $a=1$ and $b=\eta$, 
for which the linear gain coefficients are 
$g_{1}^{(1)}=4\epsilon_3^{(1)}(1/3+2\eta/T)$ 
and $g_{2}^{(1)}=4\epsilon_3^{(1)}(\eta^{2}/3+2/T)$.
We first note that $(1,\eta)$ is asymptotically stable, 
if $\eta>9/T^2$, and is unstable otherwise. That is, 
$(1,\eta)$ undergoes a bifurcation at $\eta_{bif}=9/T^2$. 
The off-on and on-off switching are based on this bifurcation, 
and are realized dynamically by appropriate 
changes in linear amplifier gain.  
To explain this, we denote by $\eta_{th}$ the value of the decision level, 
distinguishing between on and off transmission states. 
The off-on switching is achieved by 
a fast increase in $\eta$ from $\eta_{i}<\eta_{bif}$ 
to $\eta_{f}>\eta_{bif}$, such that the steady state 
$(1,\eta)$ turns from unstable to asymptotically stable. 
Consequently, before switching, $\eta_{1}$ and $\eta_{2}$ 
tend to $\eta_{s1}>\eta_{th}$ and $\eta_{s2}<\eta_{th}$, 
while after switching, $\eta_{1}$ and $\eta_{2}$ 
tend to $1$ and $\eta>\eta_{th}$. 
Thus, transmission of sequence 2 is turned on in this case. 
On-off switching is realized in a similar manner by  
a fast decrease in $\eta$ from $\eta_{i}>\eta_{bif}$ 
to $\eta_{f}<\eta_{bif}$. In this case  
$\eta_{1}$ and $\eta_{2}$ tend to $1$ and $\eta>\eta_{th}$ 
before the switching, and to $\eta_{s1}>\eta_{th}$ 
and $\eta_{s2}<\eta_{th}$ after switching.   
As a result, transmission of sequence 2 
is turned off by the change in $\eta$.

Our coupled-NLS simulations show that stable 
ultra-long-distance transmission requires $T$ values 
larger than 15. Indeed, for smaller $T$ values, high-order effects 
that are neglected by Eqs. (\ref{Hybrid3}) and (\ref{Hybrid4}), 
such as intrasequence interaction and radiation emission, lead to pulse pattern degradation 
and to breakdown of the LV model description at large distances.   
To enable comparison with results of Ref. \cite{CPJ2013} we choose $T=20$, 
but emphasize that similar results are obtained for other T values satisfying 
$T>15$. For $T=20$, bifurcation occurs at $\eta_{bif}=0.0225$. 
In transmission stabilization and off-on switching 
we use $\eta=1>\eta_{bif}$, a choice corresponding 
to typical amplitude setups in many soliton-based  
transmission systems \cite{Mollenauer2006,Iannone98}. 
In on-off switching, we use $\eta=0.02<\eta_{bif}$. 
Note that the small $\eta$ value here is dictated by the 
small value of $\eta_{bif}$.

Let us describe in some detail stability and bifurcation analysis 
for the equilibrium states of Eq. (\ref{Hybrid3}), 
for parameter values $a=1$, $b=\eta$, and $T>15$,  
which are used in both transmission stabilization and switching \cite{Bifurcation}. 
For these parameter values, the system (\ref{Hybrid3}) 
can have  up to five steady states, located at $(1,\eta)$, $(0,0)$, 
$(A_{\eta},0)$, $(0,B_{\eta})$, and $(C_{\eta}, D_{\eta})$, 
where $A_{\eta}=\left( {1 + 6\eta /T} \right)^{1/2}$, 
$B_{\eta}=\left({\eta ^2  + 6/T} \right)^{1/2}$,  
$$C_{\eta}= \left[ {\frac{{ - q\left( {\eta} \right)}}{2} + \left( {\frac{{q^{2}\left( {\eta} \right) }}{4} + \frac{{p^{3}\left( {\eta}\right) }}{{27}}} \right)^{\frac{1}{2}} } \right]^{\frac{1}{3}}  + 
\left[ {\frac{{ - q\left( {\eta} \right)}}{2} - \left( {\frac{{q^{2}\left( {\eta} \right) }}{4} + 
\frac{{p^{3}\left( {\eta} \right) }}{{27}}} \right)^{\frac{1}{2}} } \right]^{\frac{1}{3}}  - 
\frac{1}{3},$$
$D_{\eta}= \eta+T\left( {1 - C_{\eta}^2 } \right)/6$, 
$p(\eta ) =  - 12\eta /T - 4/3$, 
and $q(\eta ) =  - 16/27 - 8\eta /T + 216/T^3$. 
Note that the first four equilibrium states exist for any $\eta>0$ and $T>0$. 
For $T>15$, the equilibrium state $(C_{\eta}, D_{\eta})$  exists provided that  
$h_{1}(\eta)>0$, where $h_{1}(\eta)=(1+6\eta/T)^{1/2}-C_{\eta}$. 
As mentioned earlier, the state $(1,\eta)$ is asymptotically stable if $\eta>9/T^2$ 
and is unstable otherwise. In contrast, the state $(0,0)$ is unstable for any 
 $\eta>0$ and $T>0$. The state $(A_{\eta},0)$ is asymptotically stable if 
 $f_{1}(\eta)=(\eta T)^3/36+\eta T^2/3-6 < 0$ and is unstable otherwise,  
 while $(0,B_{\eta})$ is asymptotically stable for $f_{2}(\eta)=T^3/36+\eta T^2/3-6 < 0$ 
 and is unstable otherwise. Finally, the steady state $(C_{\eta}, D_{\eta})$ is 
 asymptotically stable if $h_{2}(\eta)>0$ and $h_{3}(\eta)<0$, where 
 $h_{2}(\eta)=(1/3+2\eta/T-2D_{\eta}/T-C_{\eta}^{2})
 (\eta^2/3+2/T-2C_{\eta}/T-D_{\eta}^{2})-4C_{\eta}D_{\eta}/T^{2}$ and 
 $h_{3}(\eta)=(1+\eta^2)/3+2(1+\eta)/T-2(C_{\eta}+D_{\eta})/T
 -(C_{\eta}^{2}+D_{\eta}^{2})$.

\begin{figure}[htbp]
\centerline{\includegraphics[width=.8\columnwidth]{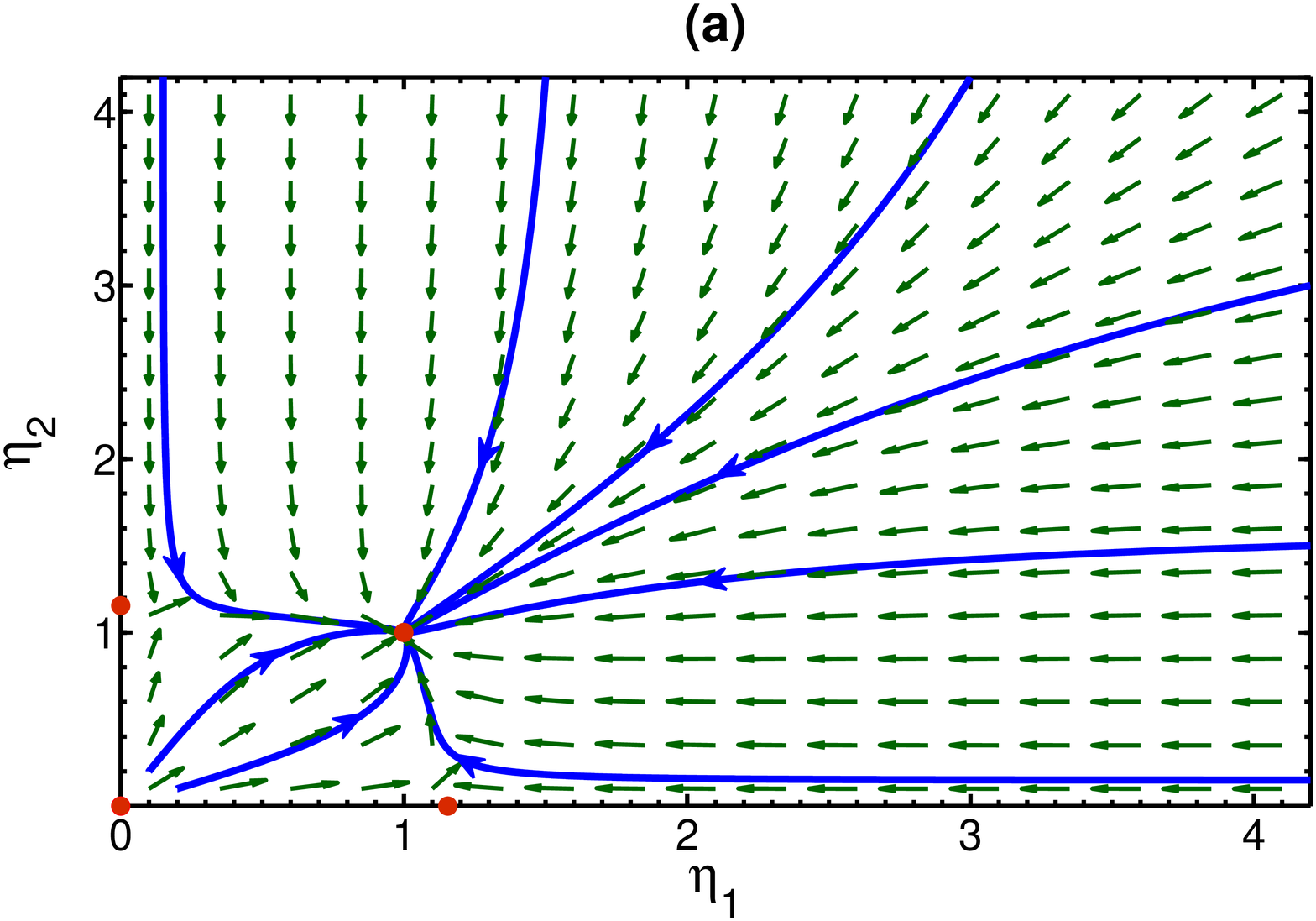}}
\centerline{\includegraphics[width=.8\columnwidth]{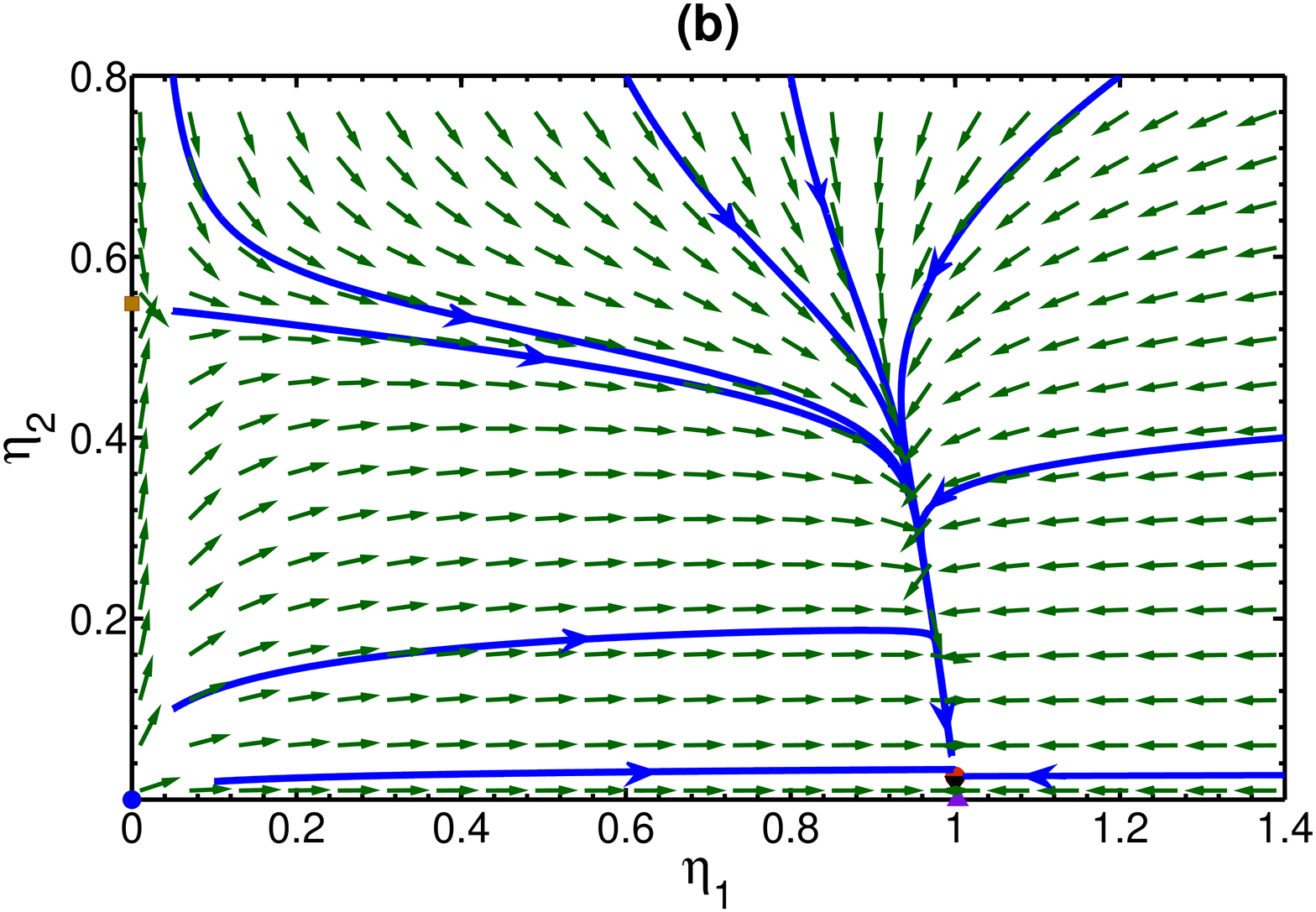}}
\caption{(Color online) 
Phase portraits for the LV model (\ref{Hybrid3}) with parameter values 
$a=1$, $b=\eta=1$, $T=20$ in (a), and $a=1$, $b=\eta=0.02$, and $T=20$ 
in (b). The blue curves are numerically calculated trajectories. 
The four red circles in (a) correspond to the four equilibrium states. 
The red circle, black down triangle, magenta up triangle, blue circle, 
and orange square in (b) represent the equilibrium states at 
$(C_{\eta},D_{\eta})$, $(1,\eta)$, $(A_{\eta},0)$,  
$(0,0)$, and $(0,B_{\eta})$, respectively, where $A_{\eta}=1.0030$, 
$B_{\eta}=0.5481$, $C_{\eta}=0.99925$, and $D_{\eta}=0.02502$.}
\label{fig1}
\end{figure}

We now describe the phase portraits of Eq. (\ref{Hybrid3}), 
for the parameter values used in our coupled-NLS simulations.  
For the set $a=1$, $b=\eta=1$, and $T=20$, 
used in transmission stabilization and off-on switching, 
Eq. (\ref{Hybrid3}) has four steady states at $(1, 1)$, 
$(0, 0)$, $(\sqrt{1.3}, 0)$, and $(0, \sqrt{1.3})$, 
of which only $(1, 1)$ is stable. In fact, as seen in the phase portrait 
of Eq. (\ref{Hybrid3}) in Fig. \ref{fig1} (a), the steady state $(1, 1)$ is 
globally asymptotically stable, i.e., the soliton amplitudes 
$\eta_{1}$ and $\eta_{2}$ both tend to $1$ for any 
nonzero input amplitudes $\eta_{1}(0)$ and $\eta_{2}(0)$.  
The global stability of the steady state $(1,1)$ is crucial 
to the robustness of pulse control in hybrid waveguide 
setups, since it allows for transmission stabilization and 
off-on switching even for input amplitude values that are 
significantly smaller or larger than 1. 
Furthermore, it can be used in broadband ``transmission recovery'', 
i.e., in the stable enhancement of pulse energies for multiple 
pulse sequences that experienced severe energy decay.

For the set $a=1$, $b=\eta=0.02$, and $T=20$, 
used in on-off switching, Eq. (\ref{Hybrid3}) has 
five steady states at  $(1,\eta)$, $(0,0)$, $(0,B_{\eta})$, 
$(A_{\eta},0)$, and $(C_{\eta},D_{\eta})$, where  
$A_{\eta}=1.0030$, $B_{\eta}=0.5481$, $C_{\eta}=0.99925$, and $D_{\eta}=0.02502$.
The first three states are unstable, while $(A_{\eta},0)$, 
and $(C_{\eta},D_{\eta})$ are asymptotically stable, as is also seen in 
the phase portrait of Eq. (\ref{Hybrid3}) in Fig. \ref{fig1} (b).   
The asymptotic stability of $(A_{\eta},0)$ and $(C_{\eta},D_{\eta})$ 
along with their proximity to $(1,0)$ enable the switching 
off of transmission of sequence 2 for a wide range of 
input amplitude values.

Note that the instability of the steady state $(0,0)$ of Eq. (\ref{Hybrid3}),  
which is related to the presence of linear gain in the waveguide, 
is a major drawback of a uniform waveguide setup with linear gain and cubic loss. 
Indeed, the presence of linear gain leads to 
enhancement of small-amplitude waves that, 
coupled with modulational instability, can cause 
severe pulse-pattern degradation. 
In the hybrid waveguide setup considered in the current paper, 
this instability is overcome by employing a GL gain-loss 
profile in even-numbered spans. 
We therefore turn to describe the results of stability analysis 
for the corresponding LV model (\ref{Hybrid4}).  
We choose $\eta=1$, and require $g_{j}^{(2)}<0$ for $j=1,2$, i.e., 
the solitons propagate in the presence of net linear loss. 
Due to the linear loss, the steady state at $(0,0)$ is asymptotically stable,   
and as a result, energies of small-amplitude 
waves decay to zero, and pulse-pattern corruption is suppressed. 
In transmission stabilization and off-on switching stabilization, 
we require that $(1,1)$ is an asymptotically stable steady state of 
Eq. (\ref{Hybrid4}). This requirement along with 
$g_{j}^{(2)}<0$ for $j=1,2$ yield the following condition \cite{PC2012}:
\begin{eqnarray}&&
(4T+90)/(5T+30)<\kappa<(8T-15)/(5T-15)
\;\;\;\; \mbox{for} \;\;\;\;  T \ge 60/17.
\label{Hybrid5a}
\end{eqnarray} 
The values $\kappa=1.6$ and $T=20$ are used in coupled-NLS simulations of transmission stabilization,   
while $\kappa=1.65$ and $T=20$ are chosen in simulations of off-on switching stabilization. 
In stabilization of on-off switching we choose $T$ and $\kappa$ values satisfying 
\begin{eqnarray}&&
 \kappa > (8T-15)/(5T-15)
\;\;\;\; \mbox{for} \;\;\;\;  T \ge 60/17,
\label{Hybrid5b}
\end{eqnarray} 
such that $(1,1)$ is unstable and another steady state at $(\eta_{s1},0)$ is asymptotically stable. 
In this manner, the switching off of soliton sequence 2 is stabilized in even-numbered spans.      
In coupled-NLS simulations for on-off switching, $\kappa=2$ and $T=20$ are used and $\eta_{s1}=1.38255$. 
We emphasize, however, that similar results are obtained for other values of $T$ and $\kappa$ 
satisfying $T>15$ and inequalities (\ref{Hybrid5a}) or (\ref{Hybrid5b}).


\section{Numerical simulations with the hybrid coupled-NLS model}
\label{Simulations}
The LV models (\ref{Hybrid3}) and (\ref{Hybrid4}) 
are based on several simplifying assumptions, whose validity might 
break down at intermediate-to-large propagation distances. 
In particular, the LV models neglect intrasequence interaction, 
radiation emission effects, and temporal inhomogeneities. 
These effects can lead to instabilities and pulse-pattern 
corruption, and also to the breakdown 
of the LV description \cite{PNC2010,PC2012}. 
In contrast, the coupled-NLS model (\ref{Hybrid1}) provides 
the full description of the propagation, which includes 
all these effects. Thus, in order to check whether 
long-distance transmission and robust broadband switching 
can be realized, it is important to carry out numerical 
simulations with the full coupled-NLS model.

The coupled-NLS system (\ref{Hybrid1}) is numerically 
solved using the split-step method with periodic boundary  
conditions \cite{Agrawal2001}. The use of periodic boundary  
conditions means that the simulations describe propagation 
in a closed waveguide loop. The initial condition consists 
of two periodic sequences of $2J+1$ overlapping solitons
with amplitudes $\eta_{j}(0)$ and zero phase:  
\begin{eqnarray} &&
\psi_{j}(t,0)\!=\!\sum_{k=-J}^{J}
\frac{\eta_{j}(0)\exp[i\beta_{j}(t-kT)]}
{\cosh[\eta_{j}(0)(t-kT)]}, 
\label{Hybrid6}
\end{eqnarray}
where $j=1,2$, and $\beta_{1}=0$, $\beta_{2}=40$, $T=20$ and 
$J=2$ are used.

\begin{figure}[htbp]
\centerline{\includegraphics[width=.8\columnwidth]{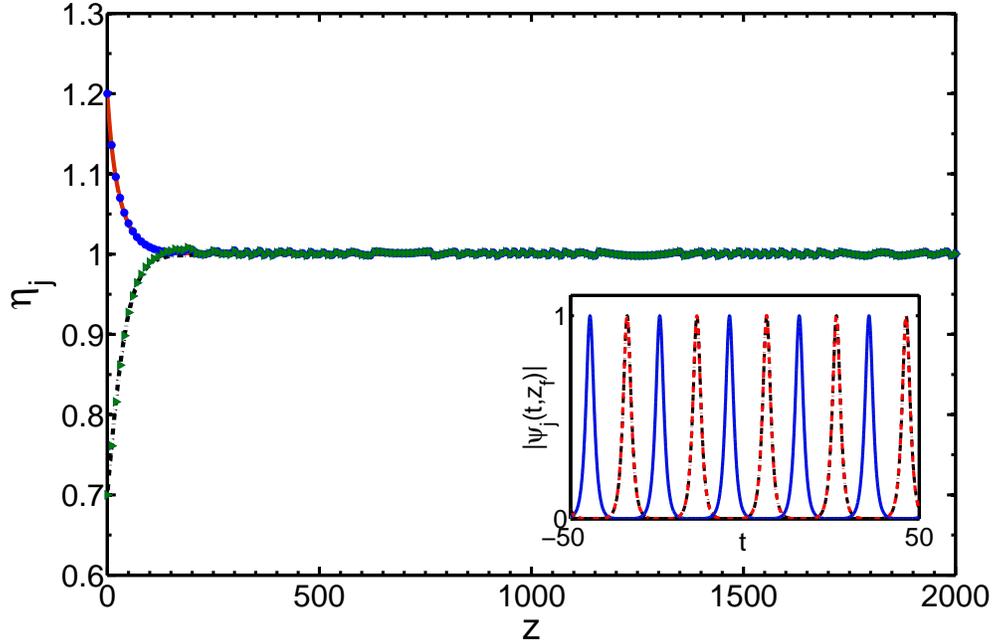}}
\caption{(Color online) The z-dependence of soliton amplitudes $\eta_{j}$ for 
transmission  stabilization with input amplitude values 
$\eta_{1}(0)=1.2$ and $\eta_{2}(0)=0.7$. 
The blue circles and green triangles represent 
$\eta_{1}(z)$ and $\eta_{2}(z)$ as obtained by numerical solution of 
Eq. (\ref{Hybrid1}), while the solid red and dashed-dotted black curves correspond to
$\eta_{1}(z)$ and $\eta_{2}(z)$ values as obtained by the LV models  
(\ref{Hybrid3}) and (\ref{Hybrid4}). 
The inset shows the final pulse patterns.  
The dashed black and solid blue lines in the inset correspond to  $|\psi_{1}(t,z_{f})|$ and 
$|\psi_{2}(t,z_{f})|$ obtained by numerical simulations with 
Eq. (\ref{Hybrid1}), while the dashed-dotted red and dotted green curves 
represent $|\psi_{1}(t,z_{f})|$ and $|\psi_{2}(t,z_{f})|$  
obtained by summation over fundamental NLS solitons with unit amplitudes, 
frequencies $\beta_{1}=0$ and $\beta_{2}=40$, and positions 
$y_{jk}(z_{f})$ for $j=1,2$ and $-2\le k \le 2$, 
which were measured from the simulations.}
\label{fig2}
\end{figure}

We first describe the results of numerical simulations for 
transmission stabilization. In this case we choose 
$a=1$ and $b=\eta=1$, so that the desired steady state of 
soliton amplitudes is $(1, 1)$. 
We use two waveguide spans $[0,150)$ and $[150,2000]$ with 
gain-loss profiles consisting of linear gain and cubic loss 
in the first span, and of linear loss, cubic gain, 
and quintic loss in the second span. The cubic loss 
coefficient in the first span is $\epsilon_{3}^{(1)}=0.015$. 
The quintic loss coefficient in the second span 
is $\epsilon_{5}=0.05$, and the ratio between cubic gain and 
quintic loss is $\kappa=\epsilon_{3}^{(2)}/\epsilon_{5}=1.6$.     
The z-dependence of $\eta_{j}$ obtained by numerical simulations 
with Eq. (\ref{Hybrid1}) for input amplitudes 
$\eta_{1}(0)=1.2$ and $\eta_{2}(0)=0.7$ is shown in Fig. \ref{fig2}. 
Also shown is the prediction of the LV models (\ref{Hybrid3}) 
and (\ref{Hybrid4}). The agreement between the coupled-NLS 
simulations and the prediction of the LV models is excellent, 
and both amplitudes tend to 1 despite 
of the fact that the input amplitude values are not close to 1. 
Furthermore, as can be seen from the inset, the shape of the soliton 
sequences is retained during the propagation. Similar results are 
obtain for other choices of input amplitude values. We emphasize that 
the distances over which stable propagation is observed are larger by 
factors of 11 and 2 compared with the distances for the uniform 
waveguide setups considered in Refs. \cite{PNC2010} and \cite{CPJ2013}. 
Additionally, the range of input amplitude values for which stable 
propagation is observed is significantly larger for hybrid waveguides 
compared with uniform ones. We therefore conclude that 
transmission stabilization is significantly enhanced by 
employing the hybrid waveguides described in the 
current paper.

We now turn to describe numerical simulations for transmission 
switching. The off-on and on-off transmission of sequence 2 is 
dynamically realized in odd-numbered spans by abrupt changes 
in the value of $\eta$ at distances $z_{s(m+1)}$ satisfying 
$z_{2m}<z_{s(m+1)}<z_{2m+1}$. These changes correspond to 
changes in the linear gain coefficients 
$g_{1}^{(1)}=4\epsilon_3^{(1)}(1/3+2\eta/T)$ 
and $g_{2}^{(1)}=4\epsilon_3^{(1)}(\eta^{2}/3+2/T)$. 
In off-on switching, $\eta=0.02$ for $z_{2m}\le z<z_{s(m+1)}$
and $\eta=1$ for $z_{s(m+1)}\le z<z_{2m+1}$, so that the steady state 
$(1,\eta)$ becomes asymptotically stable. In on-off switching, the same $\eta$ values 
are used in reverse order and $(1,\eta)$ becomes unstable. 
After switching, transmission is stabilized in even-numbered spans 
by a proper choice of $\kappa$. In off-on switching stabilization, 
$\kappa=1.65$ is used, so that $(1,1)$ is asymptotically stable. 
In on-off switching stabilization, $\kappa=2$ is used, 
so that $(1,1)$ is unstable and $(1.38255,0)$ is asymptotically stable.

The following two setups of consecutive transmission switching 
are simulated: (A) off-on-off-on-off-on-off-on, 
(B) off-on-off-on-off-on-off. We emphasize that similar results 
are obtained with other transmission switching scenarios. 
The physical parameter values in setup A are 
$T= 20$, $\epsilon_{3}^{(1)}=0.03$, $\epsilon_{5}=0.08$, 
and $\kappa_{(m+1)}=1.65$ for $0\le m \le 3$. 
The waveguide spans are determined by $z_{2m}=600m$ for 
$0\le m \le 4$ and $z_{2m+1}=140+600m$ for $0\le m \le 3$. 
That is, the spans are $[0,140)$, $[140,600)$, 
$\dots$, $[1800,1940)$, and $[1940,2400]$. 
The switching distances are $z_{s(m+1)}=100+600m$ 
for $0\le m \le 3$. 
The values of the physical parameters in setup B are 
the same as in setup A up to $z_{6}=1800$. At this distance, 
on-off switching is applied, i.e., $z_{s4}=1800$. 
In addition, $z_{7}=1940$, $z_{8}=3000$, and $\kappa=2$ 
for $z_{7}<z\le z_{8}$.

\begin{figure}[htbp]
\centerline{\includegraphics[width=.8\columnwidth]{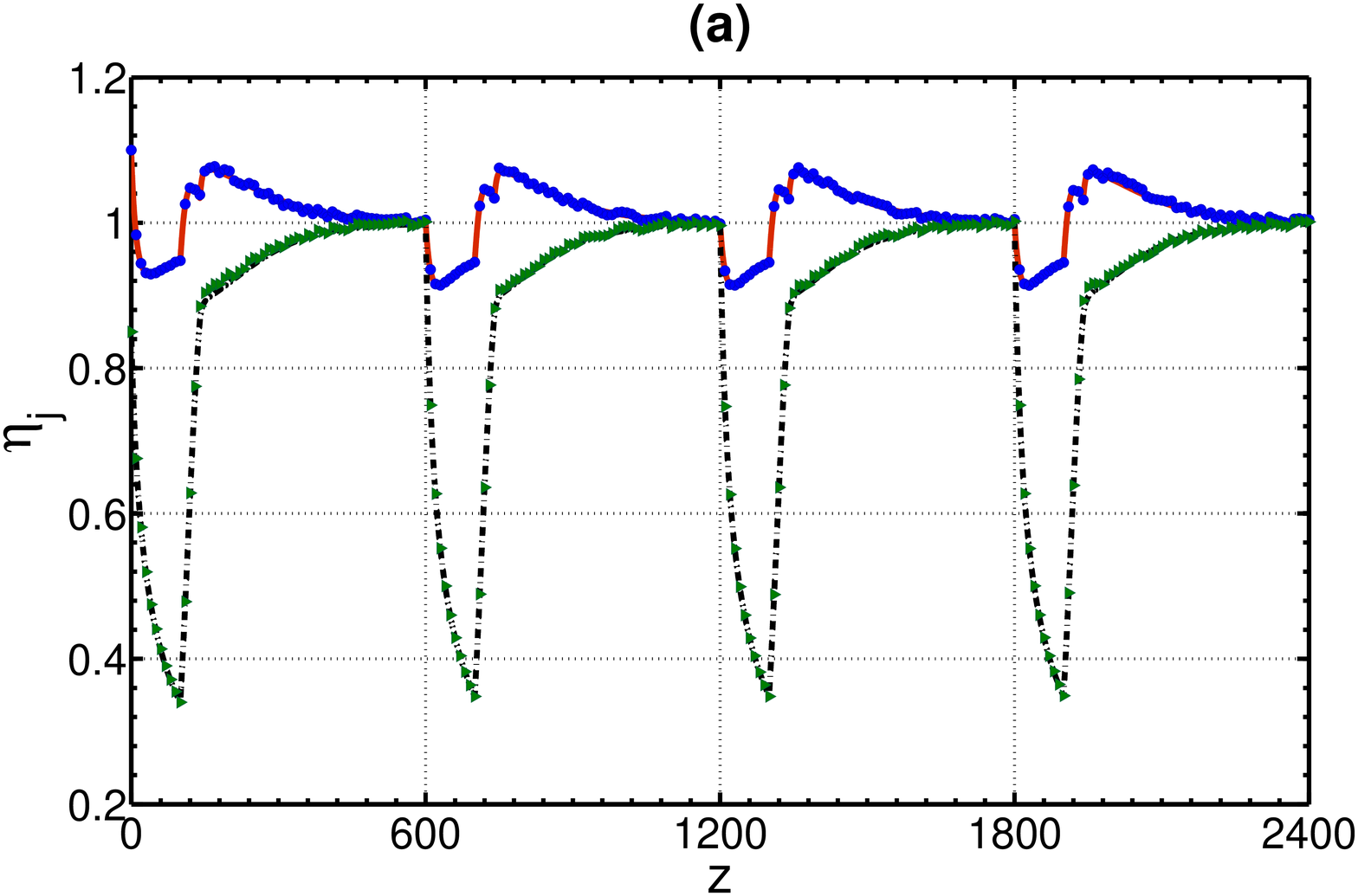}}
\centerline{\includegraphics[width=.8\columnwidth]{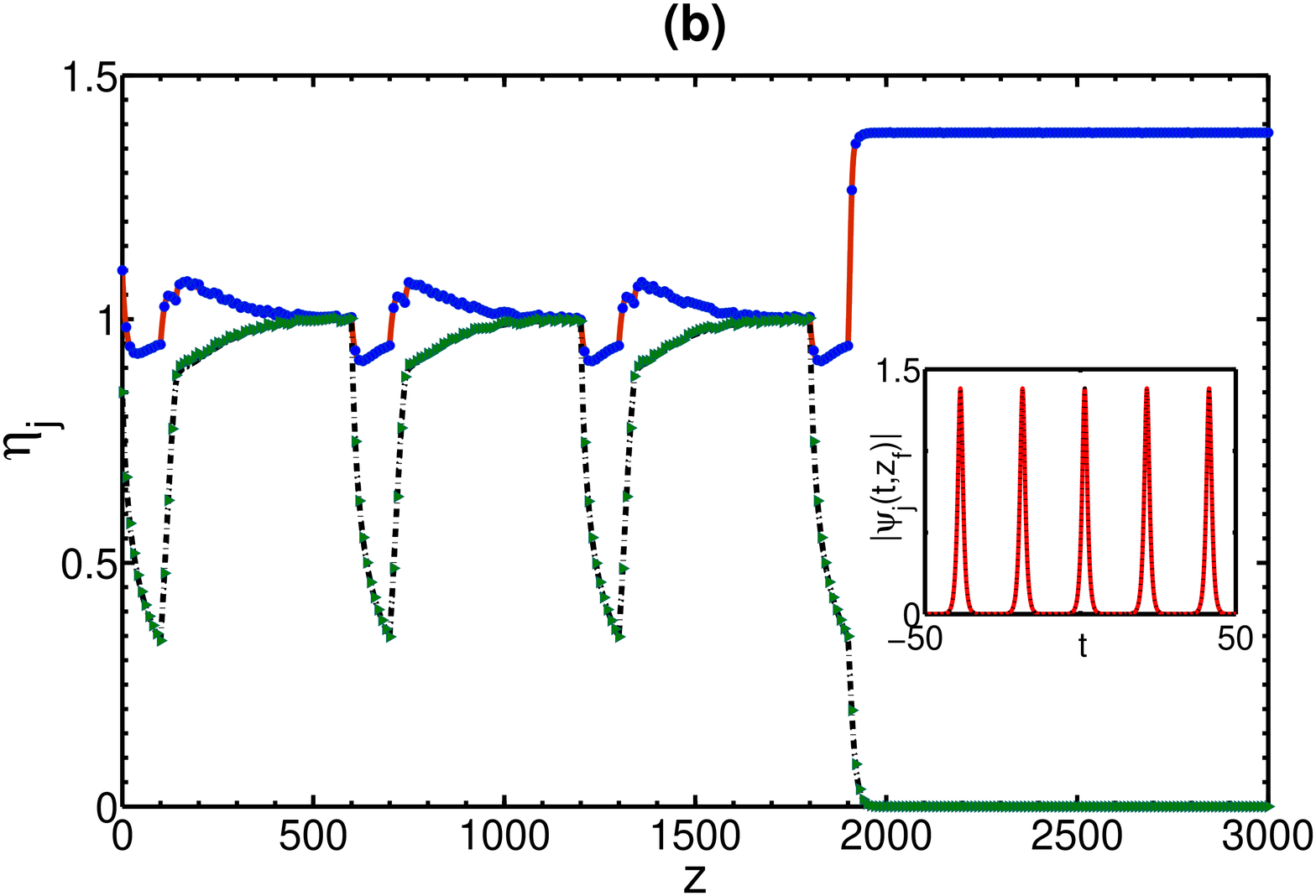}}
\caption{(Color online) The z-dependence of soliton amplitudes $\eta_{j}$ in  
multiple transmission switching setups A (a) and B (b).  
The blue circles and green triangles represent $\eta_{1}(z)$ and $\eta_{2}(z)$ as 
obtained by numerically solving Eq. (\ref{Hybrid1}), while the solid red 
and dashed-dotted black curves correspond to $\eta_{1}(z)$ and $\eta_{2}(z)$ 
predicted by the LV models (\ref{Hybrid3}) and (\ref{Hybrid4}). 
The inset shows the final pulse pattern of pulse sequence 1 
$|\psi_{1}(t,z_{f})|$ in setup B, as obtained by numerical 
solution of Eq. (\ref{Hybrid1}) (solid red curve) and by 
summation over fundamental NLS solitons (dotted black curve)
with amplitude  $\eta_{1}=1.38255$, frequency $\beta_{1}=0$, 
and positions $y_{1k}(z_{f})$ for $-2\le k \le 2$, 
which were measured from the simulations.}
\label{fig3}
\end{figure}

The results of numerical simulations with the coupled-NLS 
model (\ref{Hybrid1}) for setups A and B and input soliton 
amplitudes $\eta_{1}(0)=1.1$ and $\eta_{2}(0)=0.85$ are shown 
in Fig. \ref{fig3} (a) and (b), respectively. 
A comparison with the predictions of the LV models 
(\ref{Hybrid3}) and (\ref{Hybrid4}) is also presented. 
The agreement between the coupled-NLS simulations and 
the predictions of the LV models is excellent for both 
switching scenarios. Furthermore, as shown in the inset of Fig. \ref{fig3} (b), 
the shape of the solitons is preserved throughout the 
propagation and no growth of small amplitude waves 
(radiative instability) is observed. The propagation distances over which 
stable transmission switching is observed are larger 
by a factor of 3 compared with the distances reported in 
Ref. \cite{CPJ2013}, even though in the current paper,  
seven and eight consecutive switching events are 
demonstrated compared with only one switching event in 
Ref. \cite{CPJ2013}. Moreover, off-on transmission switching 
is observed over a large range of amplitude values 
including $\eta_{2}$ values smaller than 0.35. 
Consequently, the value of the decision level $\eta_{th}$ for 
distinguishing between on and off states can be set  
as low as $\eta_{th}=0.35$ compared with 
$\eta_{th}=0.65$ for the uniform waveguides 
considered in Ref. \cite{CPJ2013}.                   
Based on these observations we conclude that robustness 
of transmission switching is drastically increased in hybrid 
waveguide systems with a clever choice of the physical parameters. 
The increased robustness is a result of the global asymptotic stability 
of the steady state $(1,1)$ for the LV model (\ref{Hybrid3}), 
which is used to bring amplitude values close to their desired steady 
state values, and the local asymptotic stability of $(1,1)$ for the LV 
model (\ref{Hybrid4}), which is employed to stabilize 
the transmission against growth of small-amplitude waves.

\section{Conclusions}
\label{Conclusions}
In summary, we developed a method for transmission stabilization 
and switching for colliding sequences of optical solitons 
in broadband waveguide systems with nonlinear loss or gain. 
The method is based on employing hybrid waveguides, 
consisting of odd-numbered spans with linear gain and cubic 
loss, and even-numbered spans with a GL gain-loss 
profile, where the switching is dynamically realized  
by fast changes in linear amplifier gain.

We showed that dynamics of soliton amplitudes can be described 
by a hybrid LV model. Stability and bifurcation analysis of 
the steady states of the LV model was used to guide the choice 
of physical parameters values, which leads to a drastic enhancement 
in transmission stability and switching robustness. 
More specifically, the global asymptotic stability of the 
steady state $(1,1)$ of the LV model in odd-numbered spans 
was used to bring amplitude values close to their desired 
steady state values, while the local asymptotic stability of the LV model 
in even-numbered spans was employed to stabilize the transmission 
against higher-order instability due to growth of small-amplitude waves. 
Numerical simulations with the coupled-NLS equations confirmed  
the predictions of the hybrid LV model. In particular, the 
simulations showed complete suppression of radiative instability 
due to growth of small amplitude waves, which enabled  
stable propagation over distances larger by 
an order of magnitude compared with the results reported 
in Ref. \cite{PNC2010} for transmission in uniform waveguides 
with linear gain and cubic loss. Moreover, 
multiple on-off and off-on dynamic switching events, 
which are realized by fast changes in linear amplifier gain, 
were demonstrated over a wide range of soliton amplitudes, 
including amplitude values smaller than 0.35. 
As a result, the value of the decision level for 
distinguishing between on and off transmission states 
can be set as low as $\eta_{th}=0.35$, 
compared with $\eta_{th}=0.65$ for the single static 
switching event that was demonstrated in Ref. \cite{CPJ2013} 
in uniform waveguides with a GL gain-loss profile. 
Note that the increased flexibility 
in off-on switching in hybrid waveguides can be used for 
{\it transmission recovery}, i.e., for the stable amplification 
of soliton sequences, which experienced significant energy 
decay, to a desired steady state energy value.           
Based on these results, we conclude that the hybrid waveguide 
setups studied in the current paper lead to significant 
enhancement of transmission stability and switching robustness 
compared with the uniform nonlinear waveguides considered earlier.

Finally, it is worth making some remarks about potential 
applications of hybrid waveguides with different crosstalk 
mechanisms than the ones considered in the current paper. 
Of particular interest are waveguide setups, 
where the main crosstalk mechanism in odd-numbered  
and even-numbered spans are due to delayed Raman response 
and a GL gain-loss profile, respectively. 
One can envision employing these hybrid waveguides for 
enhancement of supercontinuum generation. Indeed, the interplay 
between Raman-induced energy exchange in soliton collisions and 
the Raman self-frequency shift is known to play a key role 
in widening the bandwidth of the radiation 
\cite{Frosz2006,Luan2006,Korneev2008,Dudley2011,Wabnitz2012}. 
However, the process is somewhat limited due to the fact that 
energy is always transferred from high-frequency components to 
low-frequency ones \cite{Agrawal2001}. 
This limitation can be overcome by employing waveguide spans with 
a GL gain-loss profile subsequent to spans with delayed Raman 
response. Indeed, the main effect of cubic gain on soliton 
collisions is an energy increase for both high- and low-frequency 
solitons. As a result, the energies of the high frequency 
components of the radiation will be replenished in even-numbered 
spans. This will in turn sustain the supercontinuum generation 
along longer propagation distances and might enable a wider 
radiation bandwidth compared with the one in uniform waveguides, 
where delayed Raman response is the main crosstalk-inducing mechanism.


\section*{Acknowledgments}
Q.M. Nguyen and T.P. Tran are supported by the 
Vietnam National Foundation for Science and Technology 
Development (NAFOSTED) under grant number 101.02-2012.10.


\begin{thebibliography}{}
\bibitem{Bifurcation} Note that, as described below, 
the results of our bifurcation analysis for the four equilibrium states 
$(1,\eta)$, $(0,0)$, $(A_{\eta},0)$,  and $(0,B_{\eta})$ are valid 
for any $\eta>0$ and $T>0$.
\end{thebibliography}
\end{document}